\documentclass[aps,prd,preprint,tightenlines,preprintnumbers,superscriptaddress,nofootinbib,floatfix,amsmath,amssymb]{revtex4-1}

\usepackage{graphicx}
\usepackage{hyperref}
\hypersetup{
    colorlinks=true,     
    linkcolor=blue,      
    citecolor=blue,      
    filecolor=blue,      
    urlcolor=blue        
}
\usepackage{array}
\usepackage{mathtools}
\usepackage{xspace} 

\newcommand{\Tr}{\ensuremath{\mathop{\text{Tr}}}}     

\newcommand{\rF}{{\ensuremath{F}}\xspace}
\newcommand{\rA}{{\ensuremath{A_2}}\xspace}

\newcommand{\overbar}[1]{\mkern 1.5mu\overline{\mkern-1.5mu#1\mkern-1.5mu}\mkern 1.5mu}
\newcommand{\conj}[1]{{\overbar{#1}}}

\begin{document}

\title{Stability Analysis of the Chiral Transition in SU(4) Gauge Theory with Fermions in Multiple Representations}
\author{Daniel C. Hackett}\email{daniel.hackett@colorado.edu}
\affiliation{Department of Physics, University of Colorado, Boulder, Colorado 80309, USA}

\date{December 5, 2017}

\begin{abstract}
	We present a Pisarski-Wilczek stability analysis of SU(4) gauge theory coupled simultaneously to fermions charged under the fundamental and two-index antisymmetric representations of the gauge group.
	We carry out the calculation to one loop in the $\epsilon$ expansion, assuming that the two species of fermion undergo a simultaneous chiral transition.
	The results indicate that the chiral transition is first order.
\end{abstract}


\maketitle

\section{Introduction}

Gauge theories coupled to fermions charged under multiple different irreducible representations (irreps) of the gauge group (``multirep theories'') are an old idea in Beyond Standard Model phenomenology.
In particular, multirep theories were long ago speculated to exhibit interesting phase structure via the mechanism of ``tumbling'' or ``condensation in the most attractive channel'' (MAC hypothesis) \cite{Raby:1979my}.
Phenomenologically, if observed, such a mechanism would provide a natural means of dynamically generating separated scales.
More recently, multirep theories have enjoyed a revival of interest in the context of UV completions of partially-composite Higgs models \cite{Ferretti:2013kya}.
In particular, Ferretti recently proposed such a model which contains vectorlike fermion content, and thus is amenable to investigation with lattice gauge theory \cite{Ferretti:2014qta}.
Our collaboration is carrying out an investigation of a lattice deformation of this theory, examining both the zero-temperature behavior of the theory relevant to the phenomenology of partial compositeness \cite{DeGrand:2016pgq,DeGrand:2016mxr,Ayyar:2017qdf} as well as the thermodynamics of the theory \cite{Ayyar:2017uqh,Ayyar:2017vsu,SU4FiniteT}.

With multiple irreps present in the theory, there are multiple different channels in which chiral condensation may occur.
Arguments based on tree-level gluon exchange (Casimir scaling) predict that channels involving condensation of higher irreps are more attractive, and thus condensation in these channels requires a lesser critical coupling.
Thus, in the ``tumbling'' or ``scale separation'' scenario, different channels will chirally condense at different scales as the temperature runs down from infinity, leading to interesting ``partially condensed'' phases.
Original lattice tests of this phenomenon using quenched fermions suggested it would occur \cite{Kogut:1983sm,Kogut:1982rt,Kogut:1984nq,Kogut:1984sb}, but recent lattice investigations indicate that phase separation is at least not a universal feature of vectorlike multirep theories \cite{Ayyar:2017uqh,Ayyar:2017vsu,SU4FiniteT}.

Lattice studies have determined that the lattice deformation of Ferretti's model exhibits a single phase transition between a low-temperature phase where all fermions are confined and chiral symmetry is broken, and a high-temperature phase where all fermions are deconfined and chiral symmetry is restored \cite{Ayyar:2017uqh,Ayyar:2017vsu,SU4FiniteT}.
The same investigation finds evidence that this transition is strongly first-order.
If Ferretti's model (or something similar) lies somewhere in the UV completion of the Standard Model, then the universe will have undergone this phase transition at some point early in time.
First-order transitions in the early universe produce signature gravitational waves, which may be detectable by near-future detectors \cite{Schwaller:2015tja, Caprini:2015zlo}.
If the phase transition is indeed first-order, then the presence (or lack thereof) of these signature gravitational waves can be used as probe of physics beyond the Standard Model.
Thus, it is important to provide an independent confirmation of the first-orderness of the transition in the  theory, preferably with analytics.

The rest of this paper presents a Pisarski-Wilczek stability analysis applicable to the chiral transition in both Ferretti's model and our collaboration's lattice deformation of this theory.
In Sec.~\ref{sec:setup}, we provide a sketch of the calculation to follow in the rest of the paper, and comment on the inputs to the calculation and how broadly its results may be applied.
In Sec.~\ref{sec:lagrangian}, we discuss symmetries in multirep theories and derive the EFT Lagrangian for our calculation.
In Sec.~\ref{sec:single-irrep-pw}, we review the results of the calculation for the single-irrep subsectors of the theory, which have been examined previously \cite{Pisarski:1983ms,Butti:2003nu,Basile:2004wa,Basile:2005hw} and whose fixed points span a subset of the fixed points of the full theory (see Sec.~\ref{sec:multirep-pisarski-wilczek} for discussion).
In Sec.~\ref{sec:multirep-pisarski-wilczek}, we present the results of the calculation for the full theory.
In the conclusion, we discuss the robustness of the predictions of the calculation against higher-order corrections (which are known to find fixed points in the single-irrep sectors that our treatment misses) and comment on future applications of the methods used here.

\section{Overview and Applicability}
\label{sec:setup}

We are interested in the general class of theories with a single SU($N$) gauge field coupled to fermions in one complex irrep of the gauge group and fermions in one real irrep of the gauge group.
We will explicitly consider the irrep content of Ferretti's model: $N_F$ Dirac flavors of fundamental irrep ($F$) fermions and $N^w_\rA$ Weyl flavors of two-index antisymmetric irrep (\rA) fermions in SU(4), where \rA is a real irrep.
However, much of the discussion below applies to the more general case; we will state (in Sec.~\ref{sec:lagrangian}) when our results or methods are no longer applicable this broader class of theories.
When it is not possible to state results for general $N_F$ and $N^w_\rA$, we will consider two examples:
Ferretti's model, or SU(4) gauge theory coupled to $N_F=3$ Dirac flavors of \rF fermions and $N^w_\rA=5$ Weyl flavors of \rA fermions \cite{Ferretti:2014qta};
and the ``lattice-deformed Ferretti model,'' or SU(4) gauge theory coupled to $N_F=2$ Dirac flavors of $F$ fermions and $N_\rA=2$ Dirac (or $N^w_\rA=4$ Weyl) flavors of $A_2$ fermions.

A Pisarski-Wilczek stability analysis amounts to analyzing the critical behavior of a three-dimensional effective field theory (EFT) of the two chiral condensates of our theory of interest.
Some immediate specialization is in order to steer the calculation.
We will examine only the dual-chiral limit where the fermion masses $m_F=m_\rA=0$, so that chiral symmetry is exact.
The EFT Lagrangian derived in Sec.~\ref{sec:lagrangian} is of a different form if either $N_F=1$ or $N^w_\rA=1$, so we assume $N_F>1$ and $N^w_\rA>1$ in what follows.

We also immediately specialize our analysis to the case where chiral condensation occurs simultaneously in both the $F$ and $A_2$ sectors, as is found by lattice investigations.
To probe whether simultaneous critical behavior is possible, we will simultaneously tune the condensate mass parameters for both sectors to zero.
We comment in the conclusion on how to adapt the calculation to treat the case of separated phase transitions.

We further specialize to the spontaneous chiral symmetry breaking ($\chi$SB) pattern
\begin{equation}
	\textrm{SU}(N_F)_L \times \textrm{SU}(N_F)_R \times \textrm{SU}(N^w_\rA) \times \textrm{U}(1)_A \rightarrow \textrm{SU}(N_F)_V \times \textrm{SO}(N^w_\rA)
\end{equation}
which we will motivate in Sec.~\ref{sec:lagrangian}.
Although the theory might have exhibited some other $\chi$SB pattern, this pattern is strongly supported by the consistency of lattice investigations of the zero-temperature spectroscopy of the lattice-deformed Ferretti model \cite{Ayyar:2017qdf} with chiral perturbation theory assuming this pattern \cite{DeGrand:2016pgq}.
The potential is thus subject to constraints required to induce the assumed $\chi$SB pattern, as discussed in Appendix~\ref{sec:potential}.

The procedure introduced by Pisarski and Wilczek \cite{Pisarski:1983ms} is:
first, following the usual EFT prescription, identify the symmetries of the theory and its spontaneous symmetry breaking pattern.
Then, construct the most general Landau-Ginzburg-Wilson (LGW) Lagrangian consistent with those symmetries.
Because we are interested in critical behavior, this Lagrangian includes only relevant and marginal operators.  Additionally, consider this theory in three dimensions: the finite-temperature system is compact in the (Euclidean) temporal dimension, which will trivialize as the correlation length diverges at criticality.
Compute the $\beta$ functions of the theory and identify their fixed points.
Finally, determine whether any of these fixed points are infrared-stable by examining the eigenvalues of the stability matrix $\partial\beta_{g_i} / \partial g_j$ (where $g_i$ are the couplings of the theory).
If any of the eigenvalues of $\partial\beta_{g_i} / \partial g_j$ at a fixed point are negative, that fixed point is infrared-unstable.
If no infrared-stable fixed points exist, the calculation predicts that the chiral transition must be first order.
However, if any infrared-stable fixed points exist, then the calculation predicts that the chiral transition may be second-order, provided that the transition occurs in the basin of attraction of one of the stable fixed points (but may be first-order otherwise).

We will carry out the calculation of the $\beta$ functions to one-loop order.
We are interested in the behavior of the dimensionally-reduced 3D EFT, so we employ the $\epsilon$ expansion: we expand as usual in small $\epsilon=4-d$, then set $\epsilon$ to 1.
While more sophisticated methods exist \cite{Basile:2004wa,Butti:2003nu} to treat three-dimensionality, results are scheme-independent at one loop, so the $\epsilon$ expansion provides the same results as more careful treatments.
This lowest-nontrivial-order approach is known to miss stable fixed points in cases relevant to our analysis (the $N_F=2$ single-irrep theory with suppressed anomaly) \cite{Basile:2004wa,Basile:2005hw}.
We discuss the implications for our calculation in the conclusion.

\section{Symmetries and Lagrangian}
\label{sec:lagrangian}

\subsection{Chiral Symmetry and $\chi$SB Pattern}

In order to derive the LGW Lagrangian, we must first identify the symmetries of our system.
For each irrep of fermion in a multirep theory, there is a completely independent chiral symmetry.
The fundamental irrep $F$ of SU($N_c$) is complex for $N_c > 2$, and so the $F$ sector in our theories of interest has the usual chiral symmetry
\begin{equation}
	\mathrm{SU}(N_F)_L \times \mathrm{SU}(N_F)_R
	\label{eqn:chi-symm-F}
\end{equation}
where $N_F$ is the number of Dirac flavors of $F$ fermions.
When fermions are charged under real irreps like the \rA of SU(4), chiral symmetry is enhanced by an additional $Z_2$ relating color and anticolor \cite{Peskin:1980gc,DeGrand:2015lna}.
This expands the chiral symmetry in the $A_2$ sector to
\begin{equation}
	\mathrm{SU}(N^w_\rA)
	\label{eqn:chi-symm-A2}
\end{equation}
where $N^w_\rA$ is the number of Weyl flavors of $A_2$ fermions.

Both the $F$ and $A_2$ sectors have their own independent axial $\mathrm{U}(1)$ would-be symmetries, but each of the associated axial currents is anomalous.
However, for a theory with fermions charged under $n$ different irreps, it is possible to construct $n-1$ linear combinations of the axial currents that are not anomalous (see for example Refs.~\cite{Clark:1986vk,DeGrand:2016pgq}).
Thus, for a theory with two representations of fermion, there is one non-anomalous $\mathrm{U}(1)_A$.
Taking the product of the chiral symmetry groups of each sector and the non-anomalous $\mathrm{U}(1)_A$, we find the full chiral symmetry breaking pattern of the theory is
\begin{equation}
	\textrm{SU}(N_F)_L \times \textrm{SU}(N_F)_R \times \textrm{SU}(N^w_\rA) \times \textrm{U}(1)_A \rightarrow \textrm{SU}(N_F)_V \times \textrm{SO}(N^w_\rA).
	\label{eqn:mrep-ssb-pattern}
\end{equation}

\subsection{Field Content and Single-Irrep Lagrangians}

The field content of the EFT is dictated by the chiral symmetries Eqs.~\ref{eqn:chi-symm-F} and \ref{eqn:chi-symm-A2}.
For the fundamental sector (and complex irreps in general), define the order parameter field $\phi \sim \conj{q}_{Ri} q_L^j$, where $i$ and $j$ are flavor indices. The field $\phi$ is an $N_F \times N_F$ complex matrix field which transforms under chiral rotations like
\begin{equation}
	\phi \rightarrow U_L \phi U_R^\dagger e^{2 i \alpha_F}
	\label{eqn:phi-transformation}
\end{equation}
where $U_L \in \mathrm{SU}(N_F)_L$, $U_R \in \mathrm{SU}(N_F)_R$, and $\alpha_F$ is the angle of the axial rotation in the $F$ sector.
For the antisymmetric sector (and real irreps in general), define the order parameter field $\theta \sim Q_I Q_J$, where $Q$ is a left-handed Weyl field and $I$ and $J$ are Weyl flavor indices.
The $\theta$ field is an $N^w_\rA \times N^w_\rA$ symmetric complex matrix field, which transforms under chiral rotations like
\begin{equation}
	\theta \rightarrow e^{2 i \alpha_\rA} V \theta V^T
	\label{eqn:theta-transformation}
\end{equation}
where $V \in \mathrm{SU}(N^w_\rA)$ and $\alpha_\rA$ is the axial rotation angle in the $A_2$ sector.

In the calculation's original application, Pisarski and Wilczek analyzed ``QCD'' (i.e., SU($N_c$) gauge theory for $N_c > 2$ with $N_F$ Dirac flavors of $F$-irrep fermions) \cite{Pisarski:1983ms}.
Without accounting for symmetry breaking by the axial anomaly, the most general Lagrangian invariant under Eq.~\ref{eqn:phi-transformation} and including only relevant and marginal terms is
\begin{equation}
	\mathcal{L}^\text{Single Irrep}
	= \tfrac{1}{2} \Tr \left[ \partial_\mu \phi^\dagger \partial^\mu \phi \right]
	+ r_F   \Tr \left[ \phi^\dagger \phi \right]
	+ \tfrac{1}{4} u_F  (\Tr \left[ \phi^\dagger \phi \right])^2
	+ \tfrac{1}{4} v_F   \Tr \left[ (\phi^\dagger \phi \right)^2]
	\label{eqn:L-singlerep}
\end{equation}
where the traces are over flavor.
This sub-Lagrangian governs the $F$ subsector of the full multirep theory.
For the $A_2$ sector (and for real irreps in general), the most general LGW Lagrangian consistent with Eq.~\ref{eqn:theta-transformation}, and without accounting for symmetry breaking by the axial anomaly, takes the same form \cite{Basile:2004wa}.%
\footnote{Ref.~\cite{Basile:2004wa} considered fermions charged under the adjoint irrep $G$. The adjoint irrep is always real and thus always has the chiral symmetry $\mathrm{SU}(N^w_G)$.  The chiral symmetry group is the only input to the Lagrangian, and so their results apply to all real-irrep fermions.}
Thus, the sub-Lagrangian that governs the $A_2$ subsector of the full multirep theory is Eq.~\ref{eqn:L-singlerep} with $\phi \rightarrow \theta$, $F \rightarrow \rA$.

\subsection{Constraints from Anomaly and Multirep Lagrangian}

As written, each sub-Lagrangian separately respects the independent axial symmetries $\mathrm{U}(1)_\rF$ and $\mathrm{U}(1)_\rA$ of each subsector.
To account for symmetry breaking by the axial anomaly we add terms constructed from determinants, which vary as
\begin{equation}
\begin{split}
	\det\phi &\rightarrow \det \phi \, e^{2 i N_\rF \alpha_\rF} \\
	\det\theta &\rightarrow \det \theta \, e^{2 i N^w_\rA \alpha_\rA}
\end{split}
\end{equation}
under arbitrary chiral transformations like Eqs.~\ref{eqn:phi-transformation} and \ref{eqn:theta-transformation}.
Whatever terms we add to the Lagrangian must respect the good $\mathrm{U}(1)_A$ of the theory.
The ratio of axial charges associated with the non-anomalous $\mathrm{U}(1)_A$ symmetry is \cite{DeGrand:2016pgq}:
\begin{equation}
\frac{\alpha_{F}}{\alpha_{A_2}} = - \frac{N^w_{A_2} T(A_2)}{2 N_{F} T(F)}
\label{eqn:axial-charge-ratio}
\end{equation}
where $T(r)$ is the group trace of representation $r$.
Determinant terms invariant under simultaneous axial rotations satisfying Eq.~\ref{eqn:axial-charge-ratio} will be of the general form%
\footnote{
	Without the minus sign in Eq.~\ref{eqn:axial-charge-ratio}, these terms would have the general form $(\det\phi^\dagger)^{d_{F}} (\det\theta)^{d_{A_2}} + (\text{c.c.})$.
}
\begin{equation}
(\det\phi)^{d_{F}} (\det\theta)^{d_{A_2}} + (\text{c.c.})
\label{eqn:general-det-term}
\end{equation}
where $d_\rF$ and $d_\rA$ are positive integers.
Under a general axial rotation in both sectors, these terms vary like
\begin{equation}
	(\det\phi)^{d_\rF} (\det\theta)^{d_\rA} \rightarrow \exp\left[
		2i(d_\rF \alpha_\rF N_\rF + d_\rA \alpha_\rA N^w_\rA)
	\right] (\det\phi)^{d_\rF} (\det\theta)^{d_\rA}
\label{eqn:det-transformation}
\end{equation}
which, demanding invariance under axial rotations satisfying Eq.~\ref{eqn:axial-charge-ratio}, yields the constraint
\begin{equation}
	T(A_2) d_\rF - 2 T(F) d_\rA = 0
	\label{eqn:det-condition}
\end{equation}
independent of the number of flavors of either species present.
Note that the traces in Eq.~\ref{eqn:det-condition} are the only point in the calculation where any information enters about the specific irreps under consideration beyond whether the irrep is real or complex.
This result applies generally to the case of one complex irrep ``$F$'' and one real irrep ``$A_2$''.
Specializing to the case of $F$ and $A_2$ in SU(4) where $T(\rA) = 2 T(\rF)$, we find that the good axial charge ratio is $\alpha_\rF/\alpha_\rA=-N^w_\rA/N_\rF$ and that $d_\rF=d_\rA$.
Thus, the lowest-order term we can add to our Lagrangian has $d_\rF = d_\rA = 1$.

The dimension of the operator Eq.~\ref{eqn:general-det-term} is
\begin{equation}
[(\det\phi)^{d_\rF} (\det\theta)^{d_\rA}] = N_\rF d_\rF + N^w_\rA d_\rA = N_F + N^w_\rA
\end{equation}
where the last equality is a specialization to the lowest-order $d_F = d_\rA = 1$ term.
The lowest-order term is only non-irrelevant if the condition $N_\rF + N^w_\rA \le 4$ is satisfied, which applies to neither of the two specific theories we are interested in with $(N_\rF, N^w_\rA)=(2,4)$ or $(N_\rF, N^w_\rA)=(3,5)$.
At this point we specialize to considering the case where $N_\rF + N^w_\rA > 4$, as in both of our theories of interest.
In this case, there are no non-irrelevant anomaly-implementing terms that we can add to the Lagrangian.
Therefore, both $\mathrm{U}(1)_F$ and $\mathrm{U}(1)_\rA$ are separately good symmetries of the effective field theory.
Physically, this says that the axial anomaly is not pertinent to the physics of the chiral transition.

Finally, there is one non-irrelevant irrep-coupling term that is consistent with the symmetries,
\begin{equation}
	\delta \mathcal{L}^\text{Irrep Coupling} = \tfrac{1}{2} w \Tr [\phi^\dagger \phi ] \Tr [\theta^\dagger \theta ].
\end{equation}
Compiling terms, we find that the full Lagrangian for our theory of interest is
\begin{equation}
\begin{split}
\mathcal{L}^\text{Multirep}	
& = \Tr [\partial_\mu \phi^\dagger \partial^\mu \phi ]
+ r_\rF \Tr [ \phi^\dagger \phi ]
+ \tfrac{1}{4} u_\rF ( \Tr [ \phi^\dagger \phi ] )^2
+ \tfrac{1}{4} v_\rF \Tr [ (\phi^\dagger \phi)^2 ]
\\
& + \Tr [\partial_\mu \theta^\dagger \partial^\mu \theta ]
+ r_\rA \Tr [ \theta^\dagger \theta ]
+ \tfrac{1}{4} u_\rA ( \Tr [ \theta^\dagger \theta ] )^2
+ \tfrac{1}{4} v_\rA \Tr [ (\theta^\dagger \theta)^2 ]
\\
& + \tfrac{1}{2} w \Tr [ \phi^\dagger \phi ] \Tr [ \theta^\dagger \theta ]
.
\label{eqn:L-multirep}
\end{split}
\end{equation}
For a convenient method to calculate with this Lagrangian and field content, see Appendix~\ref{sec:calculation}.

\section{Single-irrep Pisarski-Wilczek}
\label{sec:single-irrep-pw}

The Lagrangian Eq.~\ref{eqn:L-multirep} can be decomposed as a sum of the sub-Lagrangians of the single-irrep subsectors plus an irrep coupling term.
It follows that the $\beta$ functions of the theory reduce to the single-irrep $\beta$ functions plus corrections due to irrep coupling.
Thus, to set up the full multirep calculation, we will first review the results of its previous applications to the two single-irrep sectors of our theory.
The relevant results are for the single-irrep theories with suppressed anomalies (i.e., no determinant terms in the Lagrangian).

At one loop in the $\epsilon$ expansion the $\beta$ functions for a theory with $N_F$ Dirac flavors of fundamental (or complex-irrep, in general) fermions are \cite{Pisarski:1983ms}:
\begin{alignat}{4}
	&\beta_{u_\rF} &=& -u_\rF &&
	+ (N_\rF^2 + 4) u_\rF^2
	+ 4 N_\rF u_\rF v_\rF
	+ 3 v_\rF^2
	\label{eqn:beta-F-Bu}
	\\
	&\beta_{v_\rF} &=& -v_\rF &&
	+ 6 u_\rF v_\rF
	+ 2 N_\rF v_\rF^2
	\label{eqn:beta-F-Bv}
\end{alignat}
and for a theory with $N^w_\rA$ Weyl flavors of antisymmetric (or real-irrep, in general) fermions are \cite{Basile:2004wa}:
\begin{alignat}{4}
	&\beta_{u_\rA} &=& -u_\rA &&
	+ \tfrac{1}{2}({N^w_\rA}^2 + N^w_\rA + 8) u_\rA^2
	+ 2 (N^w_\rA + 1) u_\rA v_\rA
	+ \tfrac{3}{2} v_\rA^2
	\label{eqn:beta-A2-Bu}
	\\
	&\beta_{v_\rA} &=& -v_\rA &&
	+ 6 u_\rA v_\rA
	+ (N^w_\rA + \tfrac{5}{2}) v_\rA^2
	\label{eqn:beta-A2-Bv}
\end{alignat}
where we have redefined all couplings by the same overall constant to absorb uninteresting geometric factors.
Note that the coefficient of the linear term in each $\beta$ function is the classical dimension $\epsilon=4-d=1$ of each coupling constant.
The $F$-sector $\beta$ functions Eqs.~\ref{eqn:beta-F-Bu} and \ref{eqn:beta-F-Bv} have the fixed points
\begin{equation}
\begin{split}
	(u_F, v_F) &= \left(0,0\right) \\
	(u_F, v_F) &= \left(1/[4+N_F^2],0\right).
\end{split}
\label{eqn:F-fixed-points}
\end{equation}
The $A_2$-sector $\beta$ functions Eqs.~\ref{eqn:beta-A2-Bu} and \ref{eqn:beta-A2-Bv} have the fixed points
\begin{equation}
\begin{split}
	(u_\rA, v_\rA) &= (0,0) \\
	(u_\rA, v_\rA) &= \left(2/[8+N^w_\rA + {N^w_\rA}^2],0\right).
\end{split}
\label{eqn:A2-fixed-points}
\end{equation}
The trivial fixed points are always unstable, and the calculation finds that the nontrivial fixed points are unstable for $N_F \ge 2$ and $N^w_\rA \ge 2$, respectively.
A higher-order calculation finds that there are stable fixed points in the $N_F=2$ and $N^w_\rA=2$ cases that are missed at one loop \cite{Basile:2004wa,Basile:2005hw}.
We will discuss the implications of these missed fixed points for our calculation in the conclusion.

\section{Multirep Pisarski-Wilczek}
\label{sec:multirep-pisarski-wilczek}

For the full Lagrangian Eq.~\ref{eqn:L-multirep}, we find that the $\beta$ functions are
\begin{alignat*}{4}
&\beta_{u_\rF} &=& -u_\rF &&
+ (N_\rF^2 + 4) u_\rF^2
+ 4 N_\rF u_\rF v_\rF
+ 3 v_\rF^2 
+ \tfrac{1}{2} N^w_\rA (N^w_\rA + 1) w^2
\\
&\beta_{v_\rF} &=& -v_\rF &&
+ 6 u_\rF v_\rF
+ 2 N_\rF v_\rF^2
\\
&\beta_{u_\rA} &=& -u_\rA &&
+ \tfrac{1}{2}({N^w_\rA}^2 + N^w_\rA + 8) u_\rA^2
+ 2 (N^w_\rA + 1) u_\rA v_\rA
+ \tfrac{3}{2} v_\rA^2
+ N_\rF^2 w^2
\\
&\beta_{v_\rA} &=& -v_\rA &&
+ 6 u_\rA v_\rA
+ (N^w_\rA + \tfrac{5}{2}) v_\rA^2
\\
&\beta_{w}      &=& -w &&
+ w \left[
(N_\rF^2 + 1) u_\rF
+ 2 N_\rF v_\rF
+ \tfrac{1}{2} ({N^w_\rA}^2 + N^w_\rA + 4) u_\rA
+ (N^w_\rA + 1) v_\rA
+ w
\right]
\end{alignat*}
where we have redefined all couplings by the same overall constant to absorb uninteresting geometric factors.
Comparing with the single-irrep $\beta$ functions in Sec.~\ref{sec:single-irrep-pw}, we see that $\beta_{v_F}$ and $\beta_{v_\rA}$ are unchanged.
The irrep-coupling term associated with the coupling $w$ has induced corrections to $\beta_{u_F}$ and $\beta_{u_\rA}$.
Finally, there is a completely novel $\beta_w$ associated with the irrep-coupling term.

To perform a stability analysis, we must first identify the fixed points of the $\beta$ functions.
We find six in total, which are enumerated in Tables~\ref{tab:analytic-fixed-points} and \ref{tab:sixth-fixed-point}.
Because $\beta_w$ has an overall factor of $w$, and because the term associated with the coupling $w$ is the only coupling between the $F$ and $A_2$ sectors, we may divide the fixed points in to two classes: ``decoupled product fixed points'' where $w=0$ and ``multirep fixed points'' where $w \ne 0$.

When $w=0$, the $F$ and $\rA$ sectors decouple, and so the fixed points of the full theory are simply direct products of the fixed points of each single-irrep sector discussed in Sec.~\ref{sec:single-irrep-pw}.
They are listed in the first four rows of Table~\ref{tab:analytic-fixed-points} and include the trivial fixed point.

When $w \ne 0$, the $F$ and $\rA$ sectors are coupled.
The two fixed points we find in this case are novel to the full multirep theory.
One of these has $w > 0$, while the other has $w < 0$.
The $w>0$ fixed point, listed in the last row of Table~\ref{tab:analytic-fixed-points}, can be written concisely in closed form for general $N_F$ and $N^w_\rA$.
The closed form of the $w < 0$ fixed point, while computable by computer algebra systems, is too long to be worth recording here.
In Table~\ref{tab:sixth-fixed-point}, we provide numerical values for the couplings at this fixed point for our two theories of interest.

We used numerical root finding to confirm that no fixed points were missed by our analysis.
In the region of bare parameters defined by $-10 < g_i < 10$ where ${g_i \in \{u_F, v_F, u_\rA, v_\rA, w\}}$, we find no additional fixed points for all asymptotically free $N_F > 1$ and $N_\rA > 1$.

We derive a set of constraints on the couplings in Appendix~\ref{sec:potential}, which we will summarize here.
For each irrep $r$,%
\footnote{
	When $r=F$, then $N_r = N_F$; when $r=\rA$, then $N_r = N^w_\rA$.
} vacuum stability requires that $N_r u_r + v_r > 0$ and $u_r + v_r > 0$, and for the correct chiral symmetry breaking pattern to be realized $v_r > 0$ must hold.
Requiring vacuum stability and that both irreps are chirally condensed at zero temperature yields the constraint
\begin{equation}
	0 < w < \sqrt{(u_F+v_F/N_F)(u_\rA+v_\rA/N^w_\rA)}.
	\label{eqn:zeroT-w-constraint}
\end{equation}
These conditions are strict inequalities, and so define a volume which does not include its boundary surface.
For a fixed point to be pertinent to our physics of interest, it must be accessible to the physically-interesting volume: starting with a theory inside the volume, RG flow must be able to take the theory asymptotically close to that fixed point without ever moving outside of the volume.
This is only possible for fixed points either in the interior or on the boundary of the physically-interesting volume.
The $w < 0$ multirep fixed point violates the constraint Eq.~\ref{eqn:zeroT-w-constraint}, and so is not accessible to parameter space relevant to our physics of interest and unphysical.
The remaining five fixed points sit on the boundary of the physically-interesting volume, and thus are physical.

The stability matrix $\partial \beta_{g_i} / \partial g_j$ (where $g_i \in \{u_\rF, v_\rF, u_\rA, v_\rA, w\}$) is straightforwardly computed from the $\beta$ functions and not worth reproducing here.
At each of the six fixed points, we compute the eigenvalues of the stability matrix.
We find that none of the fixed points are stable for any asymptotically free $N_F \ge 2$ and $N^w_\rA \ge 2$.
Because this is true for all six fixed points, this conclusion holds even if we ignore the constraints from Appendix~\ref{sec:potential}.
Thus, our calculation indicates that the transition should be first order for any asymptotically free theory with $N_F \ge 2$ and $N^w_\rA \ge 2$ and with no anomaly-implementing terms.

\begingroup
\setlength\extrarowheight{5pt}
\begin{table}
	\begin{ruledtabular}
	\begin{tabular}{ccc}
		$u_\rF$ & $u_\rA$ & $w$ \\ \hline
		0 & 0 & 0  \\
		$1/(4 + N_\rF^2)$ & 0 & 0 \\
		0 & $2/(8 + N^w_\rA + {N^w_\rA}^2)$ & 0 \\
		$1/(4 + N_\rF^2)$ & $2/(8 + N^w_\rA + {N^w_\rA}^2)$ & 0 \\ 
		$2/(8 + N^w_\rA + {N^w_\rA}^2 + 2 N_\rF^2)$ & $2/(8 + N^w_\rA + {N^w_\rA}^2 + 2 N_\rF^2)$ & $2/(8 + N^w_\rA + {N^w_\rA}^2 + 2 N_\rF^2)$
	\end{tabular}
	\end{ruledtabular}
	\caption{
		The five (of six total) fixed points which are amenable to concise analytic expression.
		All fixed points found have $v_\rF=v_\rA=0$.
		The first four fixed points are ``decoupled product'' fixed points, while the fifth is a fixed point novel to the multirep system.
		Couplings are in the convention of the $\beta$ functions.
	}
	\label{tab:analytic-fixed-points}
\end{table}
\endgroup

\begingroup
\setlength\extrarowheight{5pt}
\begin{table}
	\begin{ruledtabular}
		\begin{tabular}{lllccc}
			Theory & $N_\rF$ & $N_\rA$ & $u_\rF$ & $u_\rA$ & $w$ \\ \hline
			Ferretti & 3 & 5 & 0.042519 & 0.035907 & -0.035606 \\
			Lattice & 2 & 4 & 0.075832 & 0.056288 & -0.054615  \\
		\end{tabular}
	\end{ruledtabular}
	\caption{
		Values of couplings at the second multirep fixed point, which is not amenable to concise analytic expression, for our theories of interest. Again, $v_\rF = v_\rA = 0$.  
		The numerical values are computed from closed-form expressions and truncated at five significant digits.
		Couplings are in the convention of the $\beta$ functions.
	}
	\label{tab:sixth-fixed-point}
\end{table}
\endgroup

\section{Conclusions}

The analysis detailed above suggests that the simultaneous chiral transition in SU(4) gauge theory with ${N_F \ge 2}$ fundamental fermions and ${N_\rA \ge 2}$ two-index antisymmetric fermions with $N_F + N^w_\rA > 4$ must be first-order.
The results of this calculation apply more broadly, to any $\mathrm{SU}(N)$ gauge theory with fermions charged under one complex irrep and one real irrep, with a simultaneous chiral transition in both sectors, and with no non-irrelevant anomaly-implementing terms that respect the good axial symmetry.

The validity of these results depend on whether one-loop order in the $\epsilon$ expansion (i.e., lowest nontrivial order) is sufficient to exclude the existence of stable fixed points.
However, as stated in Sec.~\ref{sec:single-irrep-pw}, a more sophisticated Pisarski-Wilczek analysis of the $N_F=2$ single-irrep theory finds a stable fixed point that is missed by this one-loop $\epsilon$ expansion treatment \cite{Basile:2004wa,Basile:2005hw}.
This is of significant concern because $N_F=2$ is the fundamental flavor content of the lattice-deformed Ferretti model with $(N_F,N^w_\rA)=(2,4)$, and so this fixed point will appear in the lattice-deformed Ferreti model in higher-order versions of decoupled product fixed points with $w=0$.
However, we argue that fixed points like this one will not be stable in the multirep theory.
These same higher-order calculations find that the $A_2$ subsector is unstable for $N^w_\rA>2$ and, because $w=0$ for these fixed points, higher-order irrep-coupling corrections cannot stabilize the $A_2$ sector.
The instability of the $A_2$ sector is sufficient to render the transition first-order.
In favor of the validity of our argument, lattice investigations of the $(N_F,N^w_\rA)=(2,4)$ theory are consistent with a first-order transition \cite{Ayyar:2017uqh}.
We cannot argue against the possibility that a stable multirep fixed point with $w \ne 0$ will appear at higher orders.
Investigating this possibility would require a more sophisticated calculation, and in light of lattice results, it does not seem that any such fixed points that may exist are relevant to the transition in the $(N_F,N^w_\rA)=(2,4)$ theory.

There exist several directions for future work on this generalization of Pisarski-Wilczek.

The calculation can be adapted to treat the case of separated phases, where one chiral condensate forms before another (e.g., $r_F$ crosses through zero while $r_{A_2}$ is still positive).
In this case, there will be two transitions.  The physics of the first transition can be treated with a Lagrangian like Eq.~\ref{eqn:L-multirep}, but with only one of the condensate masses $r$ tuned to zero.  To investigate the second transition, the Lagrangian Eq.~\ref{eqn:L-multirep} must be expanded around the new ground state of the condensed species and reanalyzed.

Pisarski-Wilczek analyses take very little information about the specific irreps of the fermions.
The field content and form of the (anomaly-na\"ive) Lagrangian is determined purely by whether the irrep is complex, real, or pseudoreal.
The only point at which any further information about the irrep enters is in determining the form of any anomaly-implementing determinant terms as per the procedure used in Sec.~\ref{sec:lagrangian}.
Because there are a finite number of irreps that can be present in asymptotically free gauge theories, and because the trace of the representation only enters in determining whether any determinant terms are present, there are a finite number of multirep LGW Lagrangians.
Thus, it is tractable (if not by hand) to perform a calculation analogous to the one presented in this paper for all interesting multirep theories.

\subsection*{Acknowledgements}

Research was supported by U.S.~Department of Energy Grant Number under grant DE-SC0010005 (Colorado).  I would like to thank Tom DeGrand for suggesting this calculation, Yigal Shamir for a critical reading of this paper, and Venkitesh Ayyar, William Jay, Ethan Neil, and Ben Svetitsky for very helpful discussion.

\appendix

\section{Constraints due to Vacuum Stability and $\chi$SB Pattern}
\label{sec:potential}

The relative values of the bare couplings in Eq.~\ref{eqn:L-multirep} must be constrained to ensure that the vacuum is stable, to give the desired chiral symmetry breaking pattern, and to ensure that both irreps of the theory are chirally condensed at zero temperature.
When $r_F < 0$ the solution for the $\phi$ field corresponding to the desired spontaneous symmetry breaking (SSB) pattern
\begin{equation*}
	\mathrm{U}(N_F)_L \times \mathrm{U}(N_F)_R / \mathrm{U}(1)_V
	\rightarrow 
	\mathrm{U}(N)_V / \mathrm{U}(1)_V
\end{equation*}
is
\begin{equation}
	\phi = \phi_0 \mathbb{I},
	\label{eqn:phi-ground-state}
\end{equation}
where $I$ is the $N_F \times N_F$ identity matrix.
The Lagrangian is invariant under transformations like
\begin{equation*}
	\phi \rightarrow U_L \phi U_R^\dagger
\end{equation*}
while the ground state Eq.~\ref{eqn:phi-ground-state} is invariant under such transformations only when ${U_L = U_R = U_V}$ where $U_V \in \mathrm{U}(N_F)_V$, and so we have the correct residual symmetry for our desired SSB pattern.
Similarly, when $r_\rA < 0$ the solution for the $\theta$ field corresponding to the desired SSB pattern
\begin{equation*}
	\mathrm{U}(N^w_\rA) \rightarrow \mathrm{O}(N^w_\rA)
\end{equation*}
is
\begin{equation}
	\theta = \theta_0 \mathbb{I},
	\label{eqn:theta-ground-state}
\end{equation}
just as for $\phi$.
The Lagrangian is invariant under transformations like
\begin{equation*}
\theta \rightarrow V \theta V^T
\end{equation*}
where $V \in \mathrm{U}(N^w_\rA)$.
The ground state Eq.~\ref{eqn:theta-ground-state} is invariant under such chiral transformations only when $V V^T = \mathbb{I}$, which implies $V \in \mathrm{O}(N^w_\rA)$ as desired.

In cases where $N^w_\rA$ is even, it is possible to arrange the Weyl fermions in to Dirac fermions and demand that the theory respect a $\mathrm{U}(N^w_\rA/2)_V$ vector symmetry.
The Vafa-Witten theorem states that this symmetry will not be broken by a QCD-like theory \cite{Vafa:1983tf}.
Accommodating this condition requires us to make $\theta \propto J$, where $J$ is an $N^w_\rA \times N^w_\rA$ matrix where the diagonal $N^w_\rA/2 \times N^w_\rA/2$ blocks are zero and the off-diagonal blocks are the $N^w_\rA/2 \times N^w_\rA/2$ identity matrix \cite{Basile:2004wa}.
It is not possible to construct this matrix when $N^w_\rA$ is odd, which reflects that it is not possible to define $\mathrm{SU}(N^w_\rA/2)_V$ with an odd number of Weyl degrees of freedom \cite{DeGrand:2016pgq}.
Because $J^2 = \mathbb{I}$ and all results below are in terms of $|\theta|^2 \propto \mathbb{I}$, using $J$ in cases where $N^w_\rA$ is even would not change anything.

There is another solution for each of the $\phi$ and $\theta$ fields that corresponds to a different physically-irrelevant SSB pattern,
\begin{align}
	\phi &= \phi_0 \mathrm{I}_1  \\
	\theta &= \theta_0 \mathrm{I}_1
\end{align}
where
\begin{equation}
	(\mathrm{I}_1)_{ij} = \delta_{i \alpha} \delta_{j \alpha}
\end{equation}
where $\alpha \in [1,N^w_\rA]$ is some integer and not summed over (i.e., $\mathrm{I}_1$ is a matrix which is all zeroes except for a single 1 on the diagonal) \cite{Basile:2004wa}.
These undesired solutions provide additional channels through which the vacuum may destabilize, so they must also be taken in to account.
Even when the vacuum is stable, further constraints are required to guarantee they are not the minimum of the potential, which would give the wrong chiral symmetry breaking pattern.

Taking the product of the two possible ground states in each sector, there are four possible ground states for the overall potential.
In what follows, the couplings always appear in characteristic combinations when expressions are evaluated for a given ground state.
For notational clarity and to avoid repeatedly enumerating lengthy expressions for each of the four ground states, we will write expressions below in terms of general couplings whose definition depends on the ground state of interest.
The couplings from the single-irrep sectors are unaware of the other sector, so we define
\begin{equation}
\begin{aligned}
& R_F \equiv N_F r_F,
&&\quad U_F \equiv N_F (N_F u_F + v_F)
&\quad \text{when } \phi &\propto \mathbb{I}
\\ 
& R_F \equiv r_F,
&&\quad U_F \equiv u_F + v_F
&\quad \text{when } \phi &\propto \mathrm{I}_1
\end{aligned}
\end{equation}
and
\begin{equation}
\begin{aligned}
& R_\rA \equiv N^w_\rA r_\rA,
&& \quad U_\rA \equiv N^w_\rA (N^w_\rA u_\rA + v_\rA)
&\quad \text{when } \theta &\propto \mathbb{I}
\\ 
& R_\rA \equiv r_\rA,
&& \quad U_\rA \equiv u_\rA + v_\rA
&\quad \text{when } \theta &\propto \mathrm{I}_1.
\end{aligned}
\end{equation}
The irrep-coupling sector is aware of the ground state of both single-irrep sectors, so we define
\begin{equation}
\begin{aligned}
& W = N_F N^w_\rA w,
&\quad \text{when } (\phi, \theta) &\propto (\mathbb{I},\mathbb{I})
\\ 
& W = N_F w,
&\quad \text{when } (\phi, \theta) &\propto (\mathbb{I},\mathrm{I}_1)
\\
& W = N^w_\rA w
&\quad \text{when } (\phi, \theta) &\propto (\mathrm{I}_1, \mathbb{I})
\\
& W = w
&\quad \text{when } (\phi, \theta) &\propto (\mathrm{I}_1, \mathrm{I}_1).
\end{aligned}
\end{equation}
In the discussion that follows, one need only plug in the appropriate definitions to recover the expressions for each ground state.

Plugging in the non-trivial ground states, we find that the potential for our theory is
\begin{equation}
V(\phi_0, \theta_0)
= R_F |\phi_0|^2 + \tfrac{1}{4} U_F |\phi_0|^4 + R_\rA |\theta_0|^2 + \tfrac{1}{4} U_\rA |\theta_0|^4 + \tfrac{1}{2} W |\phi_0|^2 |\theta_0|^2.
\label{eqn:general-potential}
\end{equation}
At large values of the field, only the quartic part of the potential
\begin{equation}
V_4(|\phi_0|^2, |\theta_0|^2) = \tfrac{1}{4} U_F |\phi_0|^4 + \tfrac{1}{4} U_\rA |\theta_0|^4 + \tfrac{1}{2} W |\phi_0|^2 |\theta_0|^2
\end{equation}
is pertinent to vacuum stability.
For each of the four ground states, we require that
\begin{equation}
	\lim_{|\phi_0|^2 \rightarrow \infty}
	\lim_{|\theta_0|^2 \rightarrow \infty}
	V_4 \left( |\phi_0|^2, |\theta_0|^2 \right) > 0
	\label{eqn:stability-condition}
\end{equation}
where the inequality is strict or the full potential Eq.~\ref{eqn:general-potential} will be unbounded from below when either of $R_F < 0$ or $R_\rA < 0$.
The condition Eq.~\ref{eqn:stability-condition} must be satisfied regardless of how the two limits are taken.
To explore this condition, we demand
\begin{equation}
\begin{split}
	&\quad\lim_{|x|^2 \rightarrow \infty}
	V_4 \left( a|x|^2, b|x|^2 \right) > 0 \\
	\Leftrightarrow & \quad
	\tfrac{1}{4} U_F a^2 + \tfrac{1}{4} U_\rA b^2 + \tfrac{1}{2} W ab > 0
\end{split}
\label{eqn:multirep-ab-stability}
\end{equation}
for all $a\ge0$, $b\ge0$, $a+b>0$, and simultaneously for all four ground states.

Taking $a=0$ we find $U_F > 0$, and taking $b=0$ we find $U_\rA > 0$: the single-irrep subsectors must be independently stable.
Plugging back in for $U_F$ and $U_\rA$, we recover the stability conditions familiar from analyses of the single-rep subsectors of the multirep theory \cite{Pisarski:1983ms,Basile:2004wa,Butti:2003nu},
\begin{equation}
\begin{aligned}
N_F u_F + v_F > 0, && N^w_\rA u_\rA + v_\rA > 0, \\
u_F + v_F > 0, && u_\rA + v_\rA > 0 .
\end{aligned}
\label{eqn:singlerep-stability-conditions}
\end{equation}
Note that the $Nu+v$ conditions do not subsume the $u+v$ conditions, as $u$ may be positive or negative.
When these conditions are satisfied, it is obvious from the form of Eq.~\ref{eqn:multirep-ab-stability} that a positive $W$ cannot destabilize the potential.
To bound negative $W$s, consider a continuation of the condition Eq.~\ref{eqn:multirep-ab-stability} where we allow $a$ and $b$ to range over all real numbers including negatives.
When both of $a<0$ and $b<0$, the signs of all terms in Eq.~\ref{eqn:multirep-ab-stability} are unchanged.
However, if only one of $a$ or $b$ is negative, the sign of the $W$ term is flipped.
Thus, allowing negative $a$ and $b$ and demanding stability amounts to simultaneously requiring
\begin{equation}
	\begin{split}
		&\tfrac{1}{4} U_F a^2 + \tfrac{1}{4} U_\rA b^2 + \tfrac{1}{2} |W| ab > 0 \\
		&\tfrac{1}{4} U_F a^2 + \tfrac{1}{4} U_\rA b^2 - \tfrac{1}{2} |W| ab > 0
	\end{split}
\end{equation}
for all $a \ge 0$, $b \ge 0$, $a+b>0$.
Assuming the single-irrep subsectors are independently stable ($U>0$), the $+|W|$ subcondition will always be satisfied and the condition thus bounds only $W<0$.
With $a$ and $b$ allowed to range over all reals, condition Eq.~\ref{eqn:multirep-ab-stability} is equivalent to the requirement for a positive-definite quadratic form in $(a,b)$.
A positive-definite quadratic form has positive eigenvalues; demanding that this is true for the LHS of Eq.~\ref{eqn:multirep-ab-stability} yields the condition that $U_F U_\rA > W^2$.  Because this is only required when $W<0$, we obtain
\begin{equation}
	W > -\sqrt{U_F U_\rA}.
	\label{eqn:w-stability-condition}
\end{equation}

To ensure that the correct $\chi$SB pattern is realized, we must constrain the couplings such that the $\phi \propto \mathbb{I}$ and $\theta \propto \mathbb{I}$ solutions minimize the potential.
When $\phi$ and $\theta$ are extrema of the potential, the value of the potential can be expressed as \cite{Basile:2004wa}
\begin{equation}
	V_\text{soln.} = 
	\tfrac{1}{2} r_F |\phi|^2 +
	\tfrac{1}{2} r_\rA |\theta|^2 = \tfrac{1}{2} R_F |\phi_0|^2 + \tfrac{1}{2} R_A |\theta_0|^2.
\end{equation}
There is obviously a disordered phase where $(\phi, \theta) = (0,0)$ and thus $V=0$.
When $R_F < 0$, there exists a phase where only $\phi$ is ordered. In this phase $|\phi_0|^2 = -2R_F/U_F$ and so $V = -R_F^2/U_F$.
Similarly, $|\theta_0|^2=-2R_\rA/U_\rA$ and $V=-R_\rA^2/U_\rA$ when $R_A < 0$ and only $\theta$ is ordered.
The solution when both $\phi$ and $\theta$ are ordered is
\begin{equation}
	|\phi_0|^2 = 2 \frac{
		W R_\rA - R_F U_\rA
	}{
		U_F U_\rA - W^2
	},
	\qquad
	|\theta_0|^2 = 2 \frac{
		W R_F - R_\rA U_F
	}{
	U_F U_\rA - W^2
	}
	\label{eqn:both-irreps-soln}
\end{equation}
for which the value of the potential is
\begin{equation}
V = \frac{
	2 W R_F R_\rA - R_F^2 U_\rA - R_\rA^2 U_F
}{
U_F U_\rA - W^2
}.
\end{equation}
Because $U_F U_\rA - W^2 > 0$ for stability, the solution Eq.~\ref{eqn:both-irreps-soln} only exists when both numerators are positive, yielding the existence condition
\begin{equation}
	\frac{W}{U_F} R_F > R_\rA > \frac{U_\rA}{W} R_F.
	\label{eqn:both-reps-existence}
\end{equation}

To obtain conditions on the couplings, we demand that no ground state with $\phi \propto \mathrm{I}_1$ and/or $\theta \propto \mathrm{I}_1$ is the minimum of the potential anywhere.
For brevity, we henceforth refer to the ground state where $(\phi, \theta) \propto (\mathbb{I}, \mathbb{I})$ as the $(\mathbb{I}, \mathbb{I})$ ground state and the potential for this ground state as $V(\mathbb{I}, \mathbb{I})$, etc.
For the case where only a single sector is ordered, demanding that $V(\mathbb{I}, 0) < V(\mathrm{I}_1, 0)$ yields the condition $v_F > 0$ 
and similarly demanding that $V(0, \mathbb{I}) < V(0, \mathrm{I}_1)$ yields the condition
$v_\rA > 0$.
For the case where both $\phi$ and $\theta$ are ordered, we similarly find that $v_F>0$ ensures that both $V(\mathbb{I}, \mathbb{I}) < V(\mathrm{I}_1, \mathbb{I})$ and $V(\mathbb{I}, \mathrm{I}_1) < V(\mathrm{I}_1, \mathrm{I}_1)$  hold; and that $v_\rA>0$ ensures that both $V(\mathbb{I}, \mathbb{I}) < V(\mathbb{I}, \mathrm{I}_1)$ and  $V(\mathrm{I}_1, \mathbb{I}) < V(\mathrm{I}_1, \mathrm{I}_1)$ hold.
Applying transitivity, if $v_\rF > 0$ and $v_\rA > 0$, then the $(\mathbb{I}, \mathbb{I})$ ground state minimizes the potential when both $\phi$ and $\theta$ are ordered.
In summary, we find that we must have $v_F > 0$ and $v_\rA > 0$ to obtain the correct chiral symmetry breaking pattern in all phases.
These conditions are the same as what is found in analyses of the single-irrep subsectors of the potential \cite{Pisarski:1983ms,Butti:2003nu,Basile:2004wa}, and so irrep coupling does not seem to affect which $\chi$SB pattern is realized.

Comparing the existence condition Eq.~\ref{eqn:both-reps-existence} between different ground states, we find that when $v_F > 0$ there is a part of parameter space where the $(\mathbb{I}, \mathbb{I})$ ground state does not exist, but one or both of $(\mathrm{I}_1, \mathbb{I})$ and $(\mathrm{I}_1, \mathrm{I}_1)$ does exist.
However, we find that $V(\mathbb{I}, 0) < V(\mathrm{I}_1, \mathbb{I})$ and $V(\mathbb{I}, 0) < V(\mathrm{I}_1, \mathrm{I}_1)$ for any stable potential in these regions, so the $(\mathbb{I}, 0)$ phase continues to be the minimum until the $(\mathbb{I}, \mathbb{I})$ phase exists.
Analogous statements apply for the \rA sector.

We may impose one final physical condition: at zero temperature, we expect (and lattice data indicates \cite{Ayyar:2017uqh,Ayyar:2017vsu,SU4FiniteT}) that both irreps will be chirally broken.
This corresponds to the requirement that the $(\mathbb{I}, \mathbb{I})$ phase must exist.
Equation~\ref{eqn:both-reps-existence} indicates that the both-sectors-ordered phases only exist when
\begin{equation*}
\frac{W}{U_F} R_F > \frac{U_\rA}{W} R_F
\label{eqn:phase-existence-condition}
\end{equation*}
is satisfied.
Recalling that for both irreps $r$, $U_r>0$ for vacuum stability and $R_r < 0$ for both of $\phi$ and $\theta$ to be ordered, we obtain different conditions depending on the sign of $W$:
\begin{align}
	W^2 < U_F U_\rA &\quad \text{when } W > 0, \label{eqn:pos-w-exist-cond}\\
	W^2 > U_F U_\rA &\quad \text{when } W < 0. \label{eqn:neg-w-exist-cond}
\end{align}
The condition Eq.~\ref{eqn:neg-w-exist-cond} for $W < 0$ is incompatible with the stability condition Eq.~\ref{eqn:w-stability-condition}, and so we find that $W > 0 \Rightarrow w > 0$ for the $(\mathbb{I}, \mathbb{I})$ phase to exist.
Combining with the bound Eq.~\ref{eqn:pos-w-exist-cond} on positive $W$ and plugging in for the $(\mathbb{I}, \mathbb{I})$ ground state we find the condition
\begin{equation}
	0 < w < \sqrt{ (u_F + v_F / N_F) (u_\rA + v_\rA / N^w_\rA) }
\end{equation}
which subsumes the stability condition $W > -\sqrt{U_F U_\rA}$.

\section{Calculational Details}
\label{sec:calculation}

The chiral symmetries of the disordered phase of the multirep system determine the field content of the multirep Pisarski-Wilczek Lagrangian: an arbitrary complex $N_F \times N_F$ matrix field $\phi$, and a symmetric complex $N^w_\rA \times N^w_\rA$ matrix field $\theta$.
We may instead obtain the forms of these fields through a more physical argument: a Pisarski-Wilczek calculation may be thought of as an analysis of the scalar and pseudoscalar modes of the theory of interest.
To see this, note that the $\phi$ and $\theta$ scalar order parameter (chiral condensate) fields may be expressed in terms of the coset of broken generators $\tau^i$ and $T^i$ associated with the $\chi$SB patterns of the $F$ and $\rA$ sectors, respectively.
The axial anomaly is accounted for by symmetry-breaking terms in the Lagrangian, so the $\tau^i$ and $T^i$ each include a generator proportional to the identity.
We may decompose $\phi$ like $\phi = S P$ where $S \equiv s^j \tau^j$, with $s^j$ real, is a Hermitian matrix describing the scalar modes; and $P \equiv \exp[i p^j \tau^j]$, with $p^j$ real, is a unitary matrix describing the pseudoscalar modes.
The product $SP$ is an arbitrary complex matrix, with $2 {N_F}^2$ real degrees of freedom parameterized by the $s^j$ and $p^j$.
In this form, it is straightforward to recover the first (only non-irrelevant) term in the chiral Lagrangian by tuning couplings to decouple the scalar modes and anomalous axial pseudoscalar mode.
Similarly, for the \rA sector we may write $\theta = PSP$, where now $S$ and $P$ are also symmetric.

We may obtain a convenient basis for calculation by manipulating these physically-motivated decompositions.
By expanding the exponential in $P$, reducing products of multiple $\tau$s to sums of single $\tau$s, and gathering coefficients, we find that $\phi$ (and similarly, $\theta$) may instead be parameterized as a sum over the broken generators with complex coefficient fields.
Specifically, the field $\phi$ may be expanded in a basis of the generators of the coset
\begin{equation}
\mathrm{U}(N_F) \times \mathrm{U}(N_F) / \mathrm{U}(N_F) \approx \mathrm{U}(N_F)
\end{equation}
like
\begin{equation}
\phi^a_b = \varphi^i (\tau^i)^a_b
\end{equation}
where $\tau$ span the fundamental irrep of $\mathfrak{u}(N_F)$ and the $\phi^i$ are complex scalar fields.
If $\varphi$ is real, $\varphi^i \tau^i$ spans all Hermitian matrices, so with $\varphi$ complex, $\varphi^i \tau^i$ spans all arbitrary complex matrices.
Similarly, the field $\theta$ may be expanded in a basis of the generators of the coset
\begin{equation}
\mathrm{U}(N^w_\rA) / \mathrm{O}(N^w_\rA)
\end{equation}
like
\begin{equation}
\theta_{AB} = \vartheta^I (T^I)_{AB}
\end{equation}
where $T$ span the fundamental representation of $\mathfrak{u}(N^w_\rA) / \mathfrak{o}(N^w_\rA)$ and the $\vartheta^I$ are complex scalar fields.
The generators $T$ are Hermitian and symmetric and thus real.
If $\vartheta$ is real, $\vartheta^I T^I$ spans all symmetric real matrices, so with $\vartheta$ complex, $\vartheta^I T^I$ spans all symmetric complex matrices.

In these bases, the Feynman rules are simply those for two coupled complex $|\phi|^4$ theories with additional flavor group structure multiplying each vertex.
Computing the group-theoretic weights associated with each diagram reduces to an exercise in generator algebra.
For the coset $\mathrm{U}(N_F)$, the usual $\mathfrak{u}(N)$ algebra identities are available.
Meanwhile, the set of generators $T$ of the coset $\mathrm{U}(N^w_\rA) / \mathrm{O}(N^w_\rA)$ is not closed under commutation, so they do not form a Lie algebra and only a reduced set of generator identities is available.
Taking in to account that the generators are symmetric $T^I_{AB} = T^I_{BA}$, we find a sufficient set of identities to perform the computation is:
\begin{equation}
\begin{gathered}
\Tr \left[ T^I T^J \right] = T_F \delta^{IJ} = \delta^{IJ} \\
(T^I)_{AB} (T^I)_{CD} = \tfrac{1}{2} \left( \delta_{AC} \delta_{BD} + \delta_{AD} \delta_{BC} \right) \\
\delta^{II} = d_G^{\mathrm{U}(N)} - d_G^{\mathrm{O}(N)} = \tfrac{1}{2} N^w_\rA (N^w_\rA + 1) \\
(T^I)_{AB} (T^I)_{BC}
= \left( C_F^{\mathrm{U}(N)} - C_F^{\mathrm{O}(N)} \right) \delta_{AC}
= \tfrac{1}{2} ({N^w_\rA} + 1) \delta_{AC}
\end{gathered}
\end{equation}
where $T_F=1$ is the trace of the fundamental representations of $\mathrm{U}(N)$ and $\mathrm{O}(N)$, set to be consistent with the conventional normalization of the kinetic term for complex scalar fields, and $C_F^{\mathrm{U}(N)}$ and $C_F^{\mathrm{O}(N)}$ are the quadratic Casimirs of the fundamental representations of $\mathrm{U}(N)$ and $\mathrm{O}(N)$.
Summing all one-loop diagrams contributing to a process
and using coset generator identities to reduce the flavor group structure,
the contribution to each counterterm can be found as the coefficient of the flavor group structure associated with the corresponding coupling.

The field content of Pisarski-Wilczek Lagrangians is unchanged for theories with anomaly-implementing determinant terms, even though the symmetries are different.
Thus, we may still calculate in such theories using the 
coset expansion bases described above.
In such bases, anomaly-implementing terms expand like ${\det \phi \propto \epsilon^{ij\ldots} \varphi^i \varphi^j \ldots}$ where $\epsilon^{ij\ldots}$ is the $N_F$-index antisymmetric tensor (and similarly for $\det \theta$, $\vartheta$, $N^w_\rA$).
The number of fields $\varphi$, $\vartheta$ in such terms depends on the number of flavors in each sector, and so when such terms are present in the EFT Lagrangian, the calculation cannot be performed for general $N_F$ or $N^w_\rA$.

\bibliography{mrpw}

\end{document}